%% file: main.tex
\documentclass[10pt,conference]{IEEEtran}
\IEEEoverridecommandlockouts
\usepackage{cite}
\usepackage{amsmath,amssymb,amsfonts}
\usepackage{algorithmic}
\usepackage{graphicx}
\usepackage[hidelinks]{hyperref}
\usepackage{textcomp}
\usepackage{xcolor}
\usepackage{adjustbox}
\usepackage{flushend}
\usepackage{listings}
\usepackage{pgfplots}
\usepackage{booktabs}
\usepackage{xfrac}
\usepackage{xurl}
\usepackage{tikz}
\def\BibTeX{{\rm B\kern-.05em{\sc i\kern-.025em b}\kern-.08em
    T\kern-.1667em\lower.7ex\hbox{E}\kern-.125emX}}

\usetikzlibrary{arrows.meta,calc,positioning}

\pgfplotsset{compat=1.18}

\lstdefinestyle{mystyle}{
    numberstyle=\tiny,
    basicstyle=\ttfamily\footnotesize,
    breakatwhitespace=false,
    breaklines=true,
    captionpos=b,
    keepspaces=true,
    numbers=left,
    numbersep=5pt,
    showspaces=false,
    showstringspaces=false,
    showtabs=false,
    tabsize=4
}
\renewcommand*{\figureautorefname}{\figurename\kern-0.1667em}

\begin{document}

\title{GPU-Native Compressed Neighbor Lists\\with a Space-Filling-Curve Data Layout\\
\thanks{Swiss Platform for Advanced Scientific Computing (PASC).}
}

\author{\IEEEauthorblockN{Felix~Thaler \\}
\IEEEauthorblockA{\textit{Swiss National Supercomputing Centre (CSCS)} \\
Zürich, Switzerland \\
felix.thaler@cscs.ch \\}
\and
\IEEEauthorblockN{Sebastian~Keller \\}
\IEEEauthorblockA{\textit{Swiss National Supercomputing Centre (CSCS)} \\
Zürich, Switzerland \\
sebastian.keller@cscs.ch}
}
\maketitle%
\begin{tikzpicture}[remember picture,overlay]
    \node[anchor=south,yshift=20pt] at (current page.south) {
        \parbox{\dimexpr\textwidth-\fboxsep-\fboxrule\relax}{
            This work has been submitted to the IEEE for possible publication.
            Copyright may be transferred without notice, after which this version may no longer be accessible.
        }
    };
\end{tikzpicture}%

\input{abstract}

\input{sec1_introduction}

\input{sec2_implementation}

\input{sec3_results}

\input{sec4_evrard}

\input{sec5_limitations}

\input{sec6_conclusion}

\section*{Acknowledgment}

This work was supported by the Swiss Platform for Advanced Scientific Computing (PASC).

\bibliographystyle{IEEEtran}
\bibliography{IEEEabrv,references}

\end{document}

%% file: abstract.tex
\begin{abstract}

We have developed a compressed neighbor list for short-range particle-particle interaction based on a space-filling curve (SFC) memory layout and particle clusters.
The neighbor list can be constructed efficiently on GPUs, supporting NVIDIA and AMD hardware, and has a memory footprint of less than 4~bytes per particle to store approximately 200 neighbors.
Compared to the highly-optimized domain-specific neighbor list implementation of GROMACS, a molecular dynamics code, it has a comparable cluster overhead and delivers similar performance in a neighborhood pass.

Thanks to the SFC-based data layout and the support for varying interaction radii per particle, our neighbor list performs well for systems with high density contrasts, such as those encountered in many astrophysical and cosmological applications.
Due to the close relation between SFCs and octrees, our neighbor list seamlessly integrates with octree-based domain decomposition and multipole-based methods for long-range gravitational or electrostatic interactions.
To demonstrate the coupling between long- and short-range forces, we simulate an Evrard collapse, a standard test case for the coupling between hydrodynamical and gravitational forces, on up to 1024~GPUs, and compare our results to the analytical solution.

\end{abstract}

\begin{IEEEkeywords}
High performance computing (HPC), molecular dynamics, neighbor list, particle neighbor search, smoothed particle hydrodynamics (SPH).
\end{IEEEkeywords}

%% file: sec1_introduction.tex
\section{Introduction}

Classical data structures for nearest-neighbor searches within a fixed cutoff radius to compute particle-particle interactions are cell lists\cite{quentrec_new_1973} and Verlet lists\cite{verlet_computer_1967}.
Also, various combinations of both are possible, e.g., LAMMPS\cite{thompson_lammps_2022} uses a cell list for building a Verlet list, while pairwise Verlet lists\cite{gonnet_pairwise_2012} use cell-local Verlet lists.

Further, hash-map based methods exist, e.g., Compact Hashing \cite{ihmsen_parallel_2011} or classical Spatial Hashing\cite{teschner_optimized_2003} that was initially developed for object collision detection in computer graphics.
Currently, they all belong to the cell-based approaches, as particle hash values are based on spatial cell indices.
A major disadvantage is the hash-induced poor memory locality, which can limit scalability, especially in large-scale simulations where the required hash table and particle data does not fit into the device cache.
They are typically employed in small-scale single-node scenarios in real-time computer graphics with quasi-unlimited domain size, where they allow for linear scaling in the number of particles despite the large number of cells in the domain.
The run-time requirements of real-time applications also often limit simulations to small scales, with particle data fitting into the device cache.

While cell-based approaches are typically tied to a single global cutoff radius and are most effective on close-to-uniform particle distributions, tree-based approaches (e.g., \cite{fernandez-fernandez_fast_2022}) can support per-particle radii and irregular particle distributions in a straightforward manner.
First, the tree structure adapts to any particle density distribution, and second, by checking distances to tree nodes based on the current particle’s radius, unnecessary work can be excluded early during the traversal.
Tree-based neighbor search is frequently employed in combined SPH and gravity solvers for astrophysics, where trees are already constructed for Barnes-Hut\cite{barnes_hierarchical_1986} or fast multipole \cite{greengard_fast_1987} gravity solvers.
However, when directly traversing the tree during the neighbor pass, neighbor search scales with $\mathcal{O}(n\log n)$ in the number of particles, while cell list and Verlet list implementations might scale linearly.

All methods are typically employed in a two-stage setup: first building some data structure, and second, querying that data structure.
For most methods, parallelization (mainly domain decomposition) and vectorization of one or both of these steps is nontrivial.
Thus, with the rise of parallel and vectorized computing hardware, research has helped to optimize existing algorithms or has come up with new general strategies for calculating pair-interactions efficiently on single instruction, multiple data (SIMD) and single instruction multiple thread (SIMT) architectures.
Examples of the former typically optimize for memory and cache efficiency (e.g., \cite{gonnet_pairwise_2012, gonnet_pseudo-verlet_2013}), or architecture-specific peculiarities like warp divergence (e.g., \cite{rossinelli_-silico_2015}).
A prominent example of the latter is provided by the authors of GROMACS\cite{pronk_gromacs_2013,abraham_gromacs_2015,pall_tackling_2015}, a leading open-source MD simulation software, in the form of a general strategy for computing pairwise particle interaction on SIMD hardware with arbitrary vector width\cite{pall_flexible_2013}.

The GROMACS strategy for storing SIMD-aware neighbor lists is based on fixed-size clusters of spatially close particles with a compact bounding box.
The optimal cluster size depends on the hardware SIMD vector width, and SIMD instructions are used to compute all pairwise interactions between two clusters.
Further, the clustering reduces the size of the neighbor list, as neighbor cluster indices can be used instead of single particle indices and storing a single list per cluster is enough.
To ensure that memory loads and stores work on contiguous blocks, the particle data is sorted by cluster index, such that close-by particles within a cluster are guaranteed to be contiguous in memory\cite{pall_flexible_2013}.

Páll and Hess propose the following cell-based strategy to achieve spatial closeness within each cluster\cite{pall_flexible_2013}.
\begin{enumerate}
    \item Bin all particles according to their positions inside the cells of a regular grid on the xy-plane.
    \item Sort the particles within each xy-cell (column) by their z-position.
    \item Where required, add dummy particles to each xy-cell, to make the number of particles in each column divisible by the cluster size.
\end{enumerate}
Finally, neighboring clusters within a column and neighbor columns are searched to build a Verlet list of neighbor clusters per cluster.
This algorithm is successfully employed in GROMACS.
However, probably due to its complexity and non-time-critical nature, only a CPU implementation is available for building the clustered neighbor list.
Additionally, the dependence on a regular grid restricts the efficient use case to close to uniform particle distributions with a single global cutoff radius.

We propose a GPU-native alternative construction method, which relies on the average compactness of the Hilbert curve\cite{moon_analysis_2001} and on Cornerstone octree\cite{keller_cornerstone_2023}, enabling the efficient application of our clustered neighbor lists to particle data with highly irregular distributions and varying cutoff radii.
With a novel compression step, we significantly reduce the already excellent memory footprint of the GROMACS neighbor list, allowing to fit even more particles into device memory.

Due to its SFC particle layout, our method is naturally compatible with octree-based domain-decomposition which makes it easy to combine it with multipole methods in applications where particle-mesh methods for long-range forces are difficult to scale, such as in large astrophysical simulations.
Despite the broader applicability, our method still offers comparable performance to GROMACS and LAMMPS for van der Waals interactions.

%% file: sec2_implementation.tex
\section{Implementation}

\subsection{Neighborhood Search and Pass}

Cornerstone octree\cite{keller_cornerstone_2023} enforces Hilbert-curve based ordering of particle data.
Based on the compactness properties of the Hilbert curve, consecutive particles in memory are guaranteed to be close in space \emph{on average}\cite{moon_analysis_2001}.
Therefore, unlike Páll and Hess\cite{pall_flexible_2013}, we do not need to reorder particles.
Instead, we define clusters as consecutive packs of eight particles along the SFC.

For collecting neighbor clusters, we traverse the octree, starting at the root node.
In a first step, we check if any cluster particle overlaps with a node’s bounding box;
In a second step, for leaf nodes, we compare individual particle-particle distances for each candidate neighbor cluster.
By using depth-first traversal, we obtain a sorted neighbor list without further ado.

Similar to GROMACS\cite{pall_flexible_2013}, we only store one neighbor list per super-cluster of 64 particles (consisting of 8 consecutive clusters) and utilize bitmasks to encode effective closeness of a neighbor cluster to each of the clusters within the super-cluster.
However, we replace GROMACS’ data structure by a somewhat simpler alternative, which stores per-super-cluster neighbor data in contiguous memory and does not use particle-particle exclusion masks, as we do not need to exclude interactions between certain particle pairs in our main application (SPH) apart from correct handling of symmetric calculations in cluster self-interactions.
For these, we calculate the required exclusions on-the-fly.
Further, we implement periodic boundary condition handling without shift vectors, that is, we always compute the periodic distances between particle pairs when required.
Per super-cluster, we store the number of cluster neighbors, bitmasks (8 bits for each neighbor cluster), and finally the (optionally compressed) array of neighbor cluster indices.

\begin{figure}
    \begin{tikzpicture}
        [every node/.style={font=\footnotesize},label/.style={text width=2.6cm,align=right,inner sep=1mm},box/.style={draw=black,minimum width=6mm,minimum height=4mm,anchor=center,fill opacity=0.88},outer/.style={anchor=west,column sep=-\pgflinewidth,inner sep=0mm}]
        \definecolor{c0}{rgb} {0.48,0.46,0.76};
        \definecolor{c1}{rgb} {1.00,0.43,0.61};
        \definecolor{c2}{rgb} {0.96,0.13,0.59};
        \definecolor{c3}{rgb} {0.09,0.75,0.77};
        \definecolor{c4}{rgb} {0.95,0.56,0.49};
        \definecolor{c5}{rgb} {0.40,0.91,0.93};
        \matrix [outer] at (0, 0) {
            \node[label] (slabel) {Super-cluster info}; &
            \node[box,fill=c0] (s0) {}; &
            \node[box,fill=c0] (s1) {}; &
            \node[box,fill=c0] (s2) {}; &
            \node[box,fill=c0] (s3) {}; &
            \node[box,fill=c0] (s4) {}; &
            \node[box,fill=c0] (s5) {}; &
            \node[box,fill=c0] (s6) {}; \\
        };
        \matrix [outer] at (0, -8mm) {
            \node[label] {Shift vectors}; &
            \node[box,fill=c3] (v0) {}; &
            \node[box,fill=c3] (v1) {}; &
            \node[box,fill=c3] (v2) {}; &
            \node[box,fill=c3] (v3) {}; \\
        };
        \matrix [outer] at (0, -16mm) {
            \node[label] {Packed NB data}; &
            \node[box,fill=c1,minimum width=3mm] (p0) {}; &
            \node[box,fill=c1,minimum width=6mm] (p1) {}; &
            \node[box,fill=c1,minimum width=8mm] (p2) {}; &
            \node[box,fill=c1,minimum width=4mm] (p3) {}; &
            \node[box,fill=c1,minimum width=16mm] (p4) {}; &
            \node[box,fill=c1,minimum width=2mm] (p5) {}; &
            \node[box,fill=c1,minimum width=12mm] (p6) {}; &
            \node[box,fill=c1,minimum width=4mm] (p7) {}; &
            \node[box,fill=c1,minimum width=3mm] (p8) {}; \\
        };
        \matrix [outer] at (0, -24mm) {
            \node[label] {Exclusion masks}; &
            \node[box,fill=c4] (e0) {}; &
            \node[box,fill=c4] (e1) {}; &
            \node[box,fill=c4] (e2) {}; &
            \node[box,fill=c4] (e3) {}; &
            \node[box,fill=c4] (e4) {}; \\
        };
        \matrix [outer] at (0, -32mm) {
            \node[label] {Particle data}; &
            \node[box,fill=c2] (d0) {}; &
            \node[box,fill=c2] (d1) {}; &
            \node[box,fill=c2] (d2) {}; &
            \node[box,fill=c2] (d3) {}; &
            \node[box,fill=c2] (d4) {}; &
            \node[box,fill=c2] (d5) {}; &
            \node[box,fill=c2] (d6) {}; &
            \node[box,fill=c2] (d7) {}; &
            \node[box,fill=c2] (d8) {}; &
            \node[box,fill=c2] (d9) {}; \\
        };

        \draw[-{Latex[length=1.5mm]}] (s4) to[out=270,in=90,looseness=1.1] (p6);
        \draw[-{Latex[length=1.5mm]}] (s4) to[out=270,in=90,looseness=0.8] (v2);

        \draw[-{Latex[length=1.5mm]}] (p1) to[out=270,in=90,looseness=1] (d0);
        \draw[-{Latex[length=1.5mm]}] (p1) to[out=270,in=90,looseness=1] (d4);
        \draw[-{Latex[length=1.5mm]}] (p1) to[out=270,in=90,looseness=0.7] (e4);
    \end{tikzpicture}
    \caption{The data arrays used by GROMACS. Arrows indicate references to other data.}
    \label{fig:datastructure-gromacs}
\end{figure}
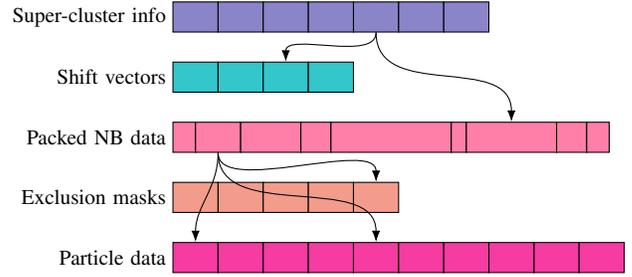

In \autoref{fig:datastructure-gromacs}, we show a visualization of the data structure used by GROMACS.
\emph{Super-cluster info} is an array storing super-cluster indices, an index pointing to a shift vector and an index pair pointing to a range in the packed neighbor data array.
The \emph{shift vectors} accommodate periodic boundary conditions.
The \emph{packed neighbor data} array contains structs of packed cluster indices, cluster interaction bitmasks (defining which clusters inside the super-cluster actually interact with a given neighbor cluster), and exclusion mask indices.
Multiple cluster indices are packed into a single struct to optimize memory loads.
The \emph{exclusion mask} array holds bitmasks that allow to exclude certain particle pairs from interacting.
This is employed for some forcing models, cluster self-interactions, and out-of-bounds particles.
\emph{Particle data} indicates the actual per-particle data (positions, forces, etc.) which, in reality, is stored in multiple arrays.

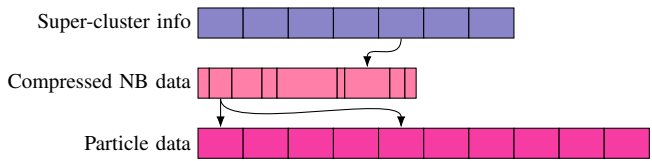
\begin{figure}
    \begin{tikzpicture}
        [every node/.style={font=\footnotesize},label/.style={text width=2.6cm,align=right,inner sep=1mm},box/.style={draw=black,minimum width=6mm,minimum height=4mm,anchor=center,fill opacity=0.88},outer/.style={anchor=west,column sep=-\pgflinewidth,inner sep=0mm}]
        \definecolor{c0}{rgb} {0.48,0.46,0.76};
        \definecolor{c1}{rgb} {1.00,0.43,0.61};
        \definecolor{c2}{rgb} {0.96,0.13,0.59};
        \definecolor{c3}{rgb} {0.09,0.75,0.77};
        \definecolor{c4}{rgb} {0.95,0.56,0.49};
        \definecolor{c5}{rgb} {0.40,0.91,0.93};
        \matrix [outer] at (0, 0) {
            \node[label] (slabel) {Super-cluster info}; &
            \node[box,fill=c0] (s0) {}; &
            \node[box,fill=c0] (s1) {}; &
            \node[box,fill=c0] (s2) {}; &
            \node[box,fill=c0] (s3) {}; &
            \node[box,fill=c0] (s4) {}; &
            \node[box,fill=c0] (s5) {}; &
            \node[box,fill=c0] (s6) {}; \\
        };
        \matrix [outer] at (0, -8mm) {
            \node[label] {Compressed NB data}; &
            \node[box,fill=c1,minimum width=1.5mm] (p0) {}; &
            \node[box,fill=c1,minimum width=3mm] (p1) {}; &
            \node[box,fill=c1,minimum width=4mm] (p2) {}; &
            \node[box,fill=c1,minimum width=2mm] (p3) {}; &
            \node[box,fill=c1,minimum width=8mm] (p4) {}; &
            \node[box,fill=c1,minimum width=1mm] (p5) {}; &
            \node[box,fill=c1,minimum width=6mm] (p6) {}; &
            \node[box,fill=c1,minimum width=2mm] (p7) {}; &
            \node[box,fill=c1,minimum width=1.5mm] (p8) {}; \\
        };
        \matrix [outer] at (0, -16mm) {
            \node[label] {Particle data}; &
            \node[box,fill=c2] (d0) {}; &
            \node[box,fill=c2] (d1) {}; &
            \node[box,fill=c2] (d2) {}; &
            \node[box,fill=c2] (d3) {}; &
            \node[box,fill=c2] (d4) {}; &
            \node[box,fill=c2] (d5) {}; &
            \node[box,fill=c2] (d6) {}; &
            \node[box,fill=c2] (d7) {}; &
            \node[box,fill=c2] (d8) {}; &
            \node[box,fill=c2] (d9) {}; \\
        };

        \draw[-{Latex[length=1.5mm]}] (s4) to[out=270,in=90,looseness=1.4] (p6);

        \draw[-{Latex[length=1.5mm]}] (p1) to[out=270,in=90,looseness=1.4] (d0);
        \draw[-{Latex[length=1.5mm]}] (p1) to[out=270,in=90,looseness=0.5] (d4);
    \end{tikzpicture}
    \caption{The data arrays used in our implementation. Arrows indicate references to other data.}
    \label{fig:datastructure-ours}
\end{figure}

Our simplified data structure is depicted in \autoref{fig:datastructure-ours}. In contrast to \autoref{fig:datastructure-gromacs}, there are no shift vectors and exclusion mask arrays, and the packed neighbor data array is replaced by a compressed data array, reducing its size. Further, GROMACS stores multiple cluster indices, interaction bitmasks, and exclusion mask indices in a packed struct inside the packed neighbor data array. Instead, we directly store all (uncompressed) bitmasks, followed by all (compressed) neighbor cluster indices for each super-cluster.

As we do not know the size of the compressed neighbor data a priori, we allocate a large array in virtual memory without populating physical pages, allowing for a hardware-assisted allocation of the required physical memory through page faults during build time.

The neighbor pass is very similar to GROMACS’ implementation.
That is, we loop in parallel over all super-clusters and process each cluster-cluster interaction of $8\times 8$~pairwise particle interactions by a single wavefront (AMD, 64~threads per wavefront) or two warps (NVIDIA, 32~threads per warp).
Before the actual computations of the interactions, if compression is enabled, we decompress the neighbor list iteratively as needed.
With compression, we typically use $8\times4$ clusters on NVIDIA GPUs, as we found them to perform slighlty better.

\subsection{Neighbor List Compression}

Our neighbor list compression incorporates ideas from general integer array compression based on variable-length-encoding (VLE) for SIMD hardware like Stream VByte\cite{lemire_stream_2018} and compression schemes optimized for neighbor lists\cite{band_compressed_2020}.
But in contrast to these algorithms, which are based on bytes as their base units, we use nibbles (half bytes, 4~bits).

Similar to \cite{band_compressed_2020}, we base our compression scheme on the observation that (sorted) neighbor lists, storing neighbor particle or cluster indices, share some common properties.
Namely, two consecutive elements in a list
\begin{enumerate}
    \item are often consecutive in value (i.e., have difference 1);
    \item otherwise, often have a rather small difference;
    \item but may have an arbitrarily large difference.
\end{enumerate}
Following these observations, and like\cite{band_compressed_2020}, we encode the differences of adjacent indices, rather than the indices themselves, employ special encoding for small numbers, especially the value 1, and use VLE to compress larger numbers.

We start by computing adjacent differences of the indices in the \emph{sorted} list of neighbors.
Then, with $w$ being the warp/workgroup size, we divide these difference values into blocks of size $w$. For each of these blocks, we then
\begin{enumerate}
    \item create a bitmask of size $w$, marking values of one;
    \item for each non-one value $v$, create a single nibble $\eta_\text{info}$, that either directly encodes the value (for values 2\,–\,9, as $\eta_\text{info}=v+6$) or otherwise encodes the number of required nibbles (values larger than 9, as $\eta_\text{info}=n_\text{nibbles}-1$);
    \item for values larger than 9, save the binary integer value as additional $n_\text{nibbles}$ data nibbles $\eta_\text{data}$.
\end{enumerate}
That is, for each value $v$ of one, we only store a single bit, for values in $[2, 9]$, we store a single bit and a nibble $\eta_\text{info}=v + 6$, and for all other values, we store a single bit, a nibble storing the length of the number in units of nibble minus one $\eta_\text{info}=n_\text{nibbles}-1$, and finally the data nibbles $\eta_\text{data}$, storing the binary value of the number itself.
For visualization purposes, \autoref{tab:nb-compression-examples} gives some examples based on a hypothetical warp size of 6.
Note that currently, we employ only a single warp for compression and decompression, which avoids inter-warp communication and allows to serially loop over all required blocks of data and simplifies packing the variable-length data.

\begin{table}
    \centering
    \caption{Compression of example difference sequences of length 6}
    \label{tab:nb-compression-examples}
    \begin{tabular}{llll}
        \toprule
        Index Differences & Bitmask & $\eta_\text{info}$ & $\eta_\text{data}$ \\
        \midrule
        1, 1, 1, 1, 1, 1 & $000000_2$ & — & — \\
        1, 2, 9, 7, 1, 1 & $011100_2$ & $\mathrm{8FD}_{16}$ & — \\
        234, 1, 1, 56789, 1, 1 & $100100_2$ & $\mathrm{13}_{16}$ & $\mathrm{EADDD5}_{16}$ \\
        \bottomrule
    \end{tabular}
\end{table}

Decompression is straightforward: We look at the bitmask, and set the output to one, where it is zero.
Then, based on the amount of non-zeros, we know the amount of info nibbles, which in turn encodes the amount of data nibbles.
Finally, we read the data either directly from $\eta_\text{info}$, respectively $\eta_\text{data}$, where required.

For a comparison to other integer encodings, \autoref{tab:nb-compression-size} shows storage requirements for the general variable-length encoding Stream VByte~\cite{lemire_stream_2018}, the neighbor list compression of~\cite{band_compressed_2020} and our compression for 32~bit integers.
Note that Stream VByte does not specialize for any specific range of numbers, as it is designed as a general integer compression scheme, while~\cite{band_compressed_2020} improves compression for values of 1 and 2.
Our algorithm further halves storage requirements for 1, uses special encoding of the values 2\,–\,9, and nibble-based VLE for the rest.

\begin{table}
    \centering
    \caption{Storage requirements of different VLE encodings}
    \label{tab:nb-compression-size}
    \begin{tabular}{lll}
        \toprule
        Encoding & Integer            & Required Bits \\
        \midrule
        Stream VByte \cite{lemire_stream_2018} & $[0, 2^8)$         & $10$       \\
                                               & $[2^8, 2^{16})$    & $18$       \\
                                               & $[2^{16}, 2^{24})$ & $26$       \\
                                               & $[2^{24}, 2^{32})$ & $34$       \\
        \midrule
        Band et al. \cite{band_compressed_2020} & $1$                 & $2$  \\
                                                & $2$                 & $2$  \\
                                                & $[3, 2^8]$          & $10$ \\
                                                & $[2^8 + 1, 2^{32}]$ & $34$ \\
        \midrule
        Ours & $1$                             & $1$        \\
             & $[2, 9]$                        & $5$        \\
             & $[10, 2^4)$                     & $9$        \\
             & $[2^{4n}, 2^{4n+4}), n\in[1,8)$ & $9+4n$     \\
        \bottomrule
    \end{tabular}
\end{table}

\subsection{Data Layout}

By considering the native warp/workgroup size, our implementation can be well optimized for specific GPU architectures.
We adapt the storage layout to the number of SIMT lanes and base the implementation on efficient warp-level scan and shuffle primitives on the GPU.
For each super-cluster requiring neighbor $j$-cluster indices, we store all data (bitmask, $\eta_\text{info}$, and $\eta_\text{data}$) interleaved in segments for groups of 32 (NVIDIA) or 64 (AMD) particles, corresponding to the hardware’s warp/workgroup size.
\autoref{tab:nb-compression-examples} shows example (bitmask, $\eta_\text{info}$, $\eta_\text{data}$) segments of length 6.

\subsection{API}

Our implementation is written in C++20 with GPU code in CUDA/HIP and employs template programming to present a simple, yet flexible and configurable interface.

\autoref{lst:api} shows a possible definition of a basic SPH density kernel which computes $\rho_i=\sum_j w_{ij}m_j$ for each particle $i$ with its neighbors $j$, with $w_{ij}$ being the SPH smoothing kernel and $m_j$ the particle mass.

Lines 2\,–\,4 show example options to configure the neighbor list at compile time; our implementation allows changing various parameters, such as cluster and super-cluster sizes, $i$\,–\,$j$ symmetry, storing only half of the neighbor pairs, and compressed and uncompressed storage.
The possibility to change these parameters provides tuning options to adapt to different hardware architectures and allows the neighbor list to accommodate different types of interaction.
Building the neighbor data structure (line 5\,–\,7) requires the locally essential tree (LET) of the Cornerstone library, the domain, total number of particles, particle group definitions (including halo information), as well as the particle coordinates and interaction radii.

\begin{lstlisting}[language=C++, label=lst:api, caption={Basic SPH density kernel implementation using our API. Required \lstinline{__device__} annotation on the lambda function was omitted for brevity.}]
using Builder = GpuSuperclusterNbListNeighborhoodBuilder<>
    ::withClusterSize<8, 8>
    ::withSuperclusterSize<64>
    ::withSymmetry
    ::withCompression<NibbleWarpCompression>;
auto neighborhood = Builder{maxNeighbors}.build(
    tree, domain, n, groups, x, y, z, h
);

neighborhood.ijLoop(
    std::tuple(m),    // input:  particle masses
    std::tuple(rho),  // output: particle densities
    [w](auto iData, auto jData, auto xij, auto dij2)
    {
        auto [i, xi, hi, mi] = iData;
        auto [j, xj, hj, mj] = jData;
        auto dij = std::sqrt(dij2);
        return std::tuple(
            symmetric::even(w(dij) * mj)
        );
    },
    ijloop::empty_postamble
);
\end{lstlisting}

Running a neighbor reduction is as simple as providing any number of input fields (line 10 in the example), output fields (line 11), the neighbor reduction itself (lines 12\,–\,18), and optionally a “postamble” function that is called after the reduction, allowing a mapping of the reduction result to the final stored value, for example, a normalization based on the neighbor particle count.

The particle pair function (lambda function or other functor), here on lines 12\,–\,18, requires particle data for $i$, $j$ particles as well as their position differences $x_{ij}$ and squared distance $d_{ij}^2$.
The passed particle data always include the particle indices, positions, and interaction radii plus additional input data; here $m_i$, respectively $m_j$, as passed on line 10.
The return value of the particle pair function has to be a tuple of output values, with one contribution for each output field; here this is just the contribution of particle $j$ to $rho_i$.
Note that the output is marked as being evenly symmetric, which allows the implementation to automatically omit double-calculation of $i$\,–\,$j$ and $j$\,–\,$i$ interactions for symmetric neighbor lists.

By default, the return values of the particle pair function are summed over all neighbor particles $j$ to compute the final value for particle $i$, but minimum and maximum reductions are also available.

Note that the particle pair function is pure (side-effect-free) and does not include any definition of the loop over the $j$-particles, enabling us to decouple pairwise particle interactions from the reduction pass over the neighborhood.
We take advantage of this separation by testing vastly different neighbor list implementations, e.g., for CPUs or a full Verlet-style list as employed by LAMMPS, with unchanged particle pair functions.
Although trivial in the case of Lennard-Jones, in SPH, these functions can become quite complex, especially in modern formulations, e.g., \cite{garcia-senz_conservative_2022}.

%% file: sec3_results.tex
\section{Results and Discussion}

\subsection{Cluster Overhead}
\label{sec:cluster-overhead}

By construction, the clustered neighbor lists lead to unnecessary calculations of particle pair interactions with distances outside the cutoff radius, because some particle pairs within an interacting cluster pair may be outside their interaction range.
However, the advantages of the approach in terms of SIMD/SIMT friendliness and storage requirements usually exceed this loss in efficiency\cite{pall_flexible_2013}.

\begin{figure}[t]
    \includegraphics{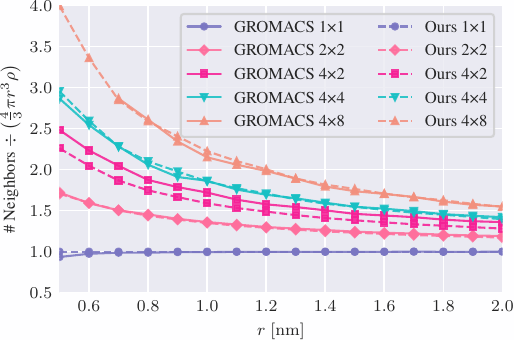}
    \caption{Cluster overhead of GROMACS’ clustering implementation, as presented in \cite[\figurename~4]{pall_flexible_2013}, vs. cluster overhead of our SFC-based approach. Number density $\rho=100\,\mathrm{nm}^{-3}$.}
    \label{fig:cluster-overhead}
\end{figure}

\autoref{fig:cluster-overhead} depicts the number of computed pair interactions in various clustered $M\times N$ pair lists, normalized by the average number of pairs in a sphere of radius $r$, as a function of the cutoff radius, as presented in \cite[\figurename~4]{pall_flexible_2013}.
Here, $M$ and $N$ refer to the target-$i$ and source-$j$ particle cluster sizes, respectively.
Evidently, the cluster overhead grows with increasing cluster size, as well as with decreasing interaction radius.
Fig.\ \ref{fig:cluster-overhead} shows that our approach leads to an almost identical cluster overhead compared to the grid-based construction algorithm of GROMACS in the uniform-density setting.
Thus, the SFC memory layout and variable particle cutoff radii that we introduced to deal with large density contrasts do not negatively impact clustering efficiency.

\subsection{Lennard-Jones Benchmark}

To demonstrate competitiveness with existing neighbor list implementations, we compare our results to the highly-optimized open-source molecular dynamics codes LAMMPS\cite{thompson_lammps_2022} and GROMACS\cite{pronk_gromacs_2013,abraham_gromacs_2015} on the NVIDIA GH200\cite{noauthor_grace_2025} and AMD MI300A\cite{noauthor_amd_2025} GPUs. LAMMPS uses full or half Verlet lists, while GROMACS uses the previously introduced clustered neighbor lists.

The benchmark consists of computing the standard 12/6 Lennard-Jones and Coulomb interactions for 4\,million atoms with close to uniform density, as is typical for large molecular dynamics simulations.
In Verlet list manner, we use a larger cutoff radius for building the neighbor list than for querying to simulate re-use of the same list for several time steps, having about 200 neighbor particles within the larger radius (i.e., neighbor list cutoff) and 150 within the smaller radius (i.e., interaction cutoff).

Besides measuring the GPU run times of the original codes of LAMMPS and GROMACS, we also ported the respective GPU kernels to our benchmark suite to get a more complete picture. We refer to these modified
kernels as LAMMPS-like and GROMACS-like.
In comparison to the original, they share the type of the neighbor list employed, but use the structure-of-array (SoA) particle data layout and SFC ordering to match our implementation.
That is, GROMACS-like also uses SFC-based clustering.
Because of a custom build step, also the order of the packed neighbor data used by the original GROMACS neighbor list and our adaption differs.
We ensured that the work per interaction pair in LAMMPS-like, GROMACS-like, and our implementation was exactly the same and added support for single, double, and mixed precision.

\begin{figure}[t]
    \centering
    \includegraphics{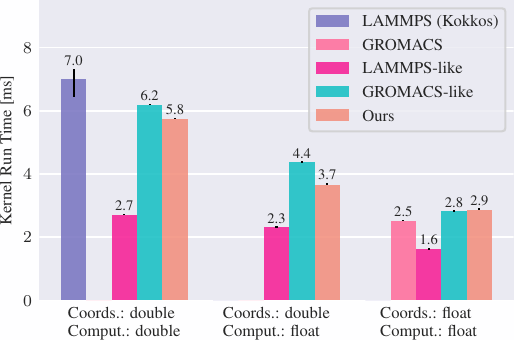}
    \caption{Performance of our pair interaction kernel for Lennard-Jones forces compared to LAMMPS and GROMACS, on the NVIDIA GH200\cite{noauthor_grace_2025}.}
    \label{fig:lj-performance-gh200}
\end{figure}

\autoref{fig:lj-performance-gh200} depicts the performance measurements on the NVIDIA GH200, the bars for LAMMPS and GROMACS show kernel run times of the original LAMMPS (Kokkos) and GROMACS codes, while LAMMPS-like and GROMACS-like show the run times of our derived kernels.
LAMMPS was compiled with its Kokkos backend in default double precision mode, GROMACS uses single precision.
Our own kernels have been measured in double precision mode, single precision mode, and mixed precision mode.
In mixed precision mode, particle coordinates are stored in double precision to retain accurate particle position differences in large-scale simulations, while pairwise interactions are computed in single precision.
The kernel run times for LAMMPS and GROMACS kernels were obtained using NVIDIA Nsight Systems, our own benchmark suite directly uses CUDA/HIP timers.
Note that original LAMMPS compared to our LAMMPS-like implementation shows an unexpectedly high run time.
This mainly arises due to additional energy computations performed within the same kernel by LAMMPS, as well as the SFC-particle ordering used in our implementation, which improved performance.
GROMACS shows a slight performance advantage compared to GROMACS-like, coming from the use of a packed array-of-structure (AoS) layout for coordinates and charge, enabling the use of fewer but wider load instructions.

\begin{figure}[t]
    \centering
    \includegraphics{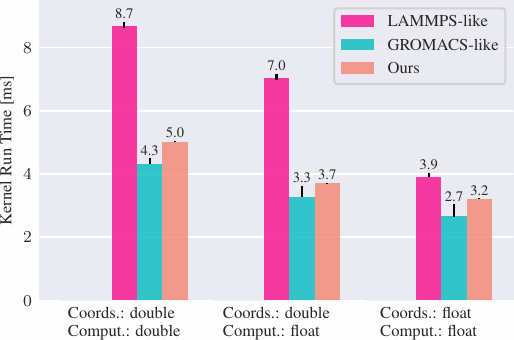}
    \caption{Performance of our pair interaction kernel for Lennard-Jones forces compared to LAMMPS and GROMACS, on the AMD MI300A\cite{noauthor_amd_2025}.}
    \label{fig:lj-performance-mi300a}
\end{figure}

On the GH200, the simple full neighbor list outperforms the clustered list in all cases in this setting with 200 neighbors, despite the orders of magnitude higher memory consumption.
In contrast, on the MI300A, the situation is the opposite:
the clustered neighbor list is significantly faster, as is visible in \autoref{fig:lj-performance-mi300a}.
We assume that this is mainly explained by the faster and especially larger caches of the NVIDIA GH200 GPU.

Overall, our clustered neighbor list provides performance similar to state-of-the-art neighbor list implementations with close to uniform particle distributions and a single global particle radius.
However, in addition, our implementation also effectively handles arbitrary particle density contrasts and per-particle cutoff radii, as demonstrated in \autoref{sec:evrard} below.

\begin{figure}[t]
    \centering
    \includegraphics{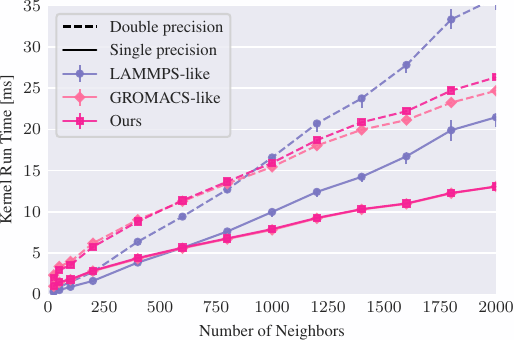}
    \caption{Performance of our pair interaction kernel for Lennard-Jones forces compared to LAMMPS and GROMACS for a wide range of neighbor counts, on the NVIDIA GH200\cite{noauthor_grace_2025}.}
    \label{fig:lj-performance-gh200-scaling}
\end{figure}

\begin{figure}[t]
    \centering
    \includegraphics{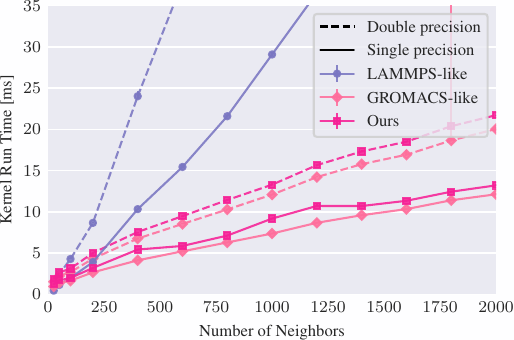}
    \caption{Performance of our pair interaction kernel for Lennard-Jones forces compared to LAMMPS and GROMACS for a wide range of neighbor counts, on the AMD MI300A\cite{noauthor_amd_2025}.}
    \label{fig:lj-performance-mi300a-scaling}
\end{figure}

In \autoref{fig:lj-performance-gh200-scaling} and \autoref{fig:lj-performance-mi300a-scaling}, we show performance results for a wide range of neighbor counts on the GH200 and MI300A.
Note that we do not show the mixed-precision results for better visual clarity.
Our neighbor lists’ performance closely matches the GROMACS neighbor list for the whole range of neighbor counts, with a slight overhead on the Mi300A for this lightweight kernel, mainly due to the required decompression.

A larger sensitivity of the MI300A to the memory footprint of the neighbor data is illustrated by the rapid performance degradation of the full neighbor list (LAMMPS-like) with an increasing number of neighbors.
Thus, on the MI300A, the clustered neighbor list almost always outperforms a full neighbor list, while on the GH200, the full neighbor list is the fastest option even for high neighbor counts—albeit with a very large memory footprint.

In the given setting, on the GH200, build times of our neighbor list range from less than 5\,ms (single-precision, approx. 25 neighbors) to 30\,ms (double precision, approx. 2000 neighbors).
In single precision, this is~1.8\,–\,4.8$\times$ the run time of the Lennard-Jones interaction kernel, and~1.1\,–\,3.1$\times$ in double precision.

On the MI300A, despite a very similar performance of the neighbor pass, we currently get much higher build times of 28\,ms to 98\,ms.
One reason besides the smaller caches is the reliance on page faults for our growing data buffer, which we measured to be significantly slower on AMD GPUs than on NVIDIA GPUs.

\subsection{Memory Footprint}

Our neighbor list implementation has a very small memory footprint.
\autoref{tab:lj-memory} shows the memory requirements in the Lennard-Jones benchmark, with approximately 200 neighbor indices saved in the neighbor list of each particle.
The full ($1\times1$) neighbor list, as used in LAMMPS, requires 800\,B per particle, which would limit the maximum feasible system size of typical SPH simulations.
GROMACS’ clustering strategy reduces the storage requirements drastically to about 12\,B per particle, while our additional compression further reduce this value by more than a factor of 3 to under 4\,B per particle.
Note that we use a cluster size of $8\times4$ on NVIDIA, but $8\times8$ on AMD GPUs.
When using compression, we found this to give the best performance on both GPUs, however, the storage requirements are slightly higher on NVIDIA, especially in the uncompressed case.
Note that even with $8\times8$ clusters, some additional memory is required on NVIDIA GPUs as two warps cooperate on a single cluster-cluster interaction and bitmasks are stored per warp (this is also visible in the GROMACS-like implementation).
In this specific case, compared to an optimal classical Verlet list, our compressed clustered neighbor list reduces the memory footprint by more than a factor of 200.

\begin{table}
\centering
\caption{Memory usage of data structures with 200 neighbors}
\begin{tabular}{lll}
\toprule
Neighbor List & GH200 & MI300A \\
\midrule
LAMMPS-like  & \phantom{0}0.8\,KB & {0.8}\,KB \\
GROMACS-like & 12.2\,B & 9.2\,B \\
Ours without Compression & 12.7\,B & 7.6\,B \\
Ours with \cite{band_compressed_2020} & \phantom{0}4.0\,B & 2.7\,B \\
Ours         & \phantom{0}3.6\,B & 2.4\,B \\
\bottomrule
\end{tabular}
\label{tab:lj-memory}
\end{table}

\begin{figure}[t]
    \centering
    \includegraphics{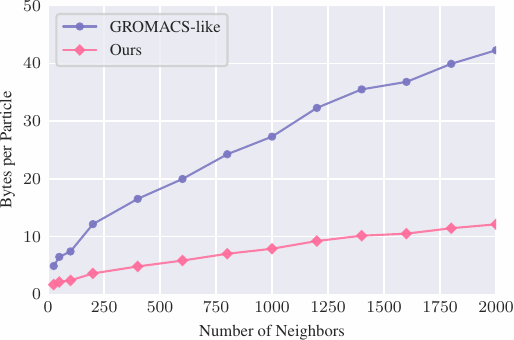}
    \caption{Memory footprint per particle of our pair interaction kernel compared to GROMACS for a wide range of neighbor counts.}
    \label{fig:lj-performance-gh200-bytes}
\end{figure}

\autoref{fig:lj-performance-gh200-bytes} shows the scaling behavior of our and GROMACS’ neighbor lists with varying neighbor counts.
Our implementation can store approximately 10 times as many neighbors compared to GROMACS with the same memory footprint.

\subsection{Compression Overhead}

Compression and decompression of the neighbor list data incur an almost negligible additional cost.
On the one hand, at build time, the reduced number of page faults when filling the neighbor data array, may actually \emph{reduce} build time significantly.
In the above Lennard-Jones benchmarks, the neighbor list build time is reduced by about 5\%\,–\,50\% when compression is enabled, depending on the hardware.
This performance benefit, however, is only significant if the neighborhood is rebuilt often, which might not be the case in real-world scenarios and is likely to depend on the specific problem (octree structure, number of neighbors).
On the other hand, decompression of the neighbor data causes a small overhead on the actual neighbor pass kernels, although this is typically a minor cost of only 1\%\,–\,5\% even for short kernels like the Lennard-Jones interaction, and negligible in larger computations.
Only in specific cases on AMD GPUs, we measured overheads of up to 20\%.

%% file: sec4_evrard.tex
\section{Evrard-Collapse Benchmark}
\label{sec:evrard}

The Evrard collapse\cite{evrard_beyond_1988} is a standard hydrodynamical test case that includes gravitational interactions between the particles.
The initial conditions consist of a gas sphere at rest with particle density proportional to $\sfrac{1}{r}$, where $r$ is the distance from the center of the sphere. Due to the lack
of initial thermal energy, gravity causes the gas sphere to contract, heating it up in the process,
which in turn gives rise to a shock wave extending outward.
We performed this test to demonstrate that the coupling of SPH with our neighbor list with a Barnes-Hut
gravity solver works correctly and to study performance and scaling in
this setting, which features large density differences and variable particle radii.

\begin{figure}[t]
    \centering
    \includegraphics{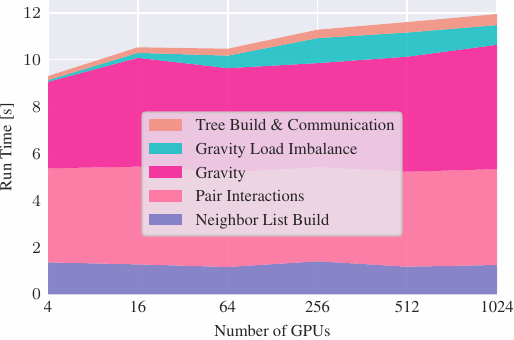}
    \caption{Weak scaling of the Evrard collapse test case on up to 1024 GH200 GPUs with 4~GPUs per node and about 250~million particles per GPU. Average time of 20 time steps is shown.}
    \label{fig:evrard-scaling-gh200}
\end{figure}

For this experiment, we have integrated our neighbor list in SPH-EXA\cite{cavelan_smoothed_2020,cabezon_sph-exa_2026}, an open-source SPH and self-gravity code and performed weak-scaling tests on the ALPS supercomputer at the Swiss National Supercomputing Centre (CSCS) on up to 1024 NVIDIA GH200 GPUs.
\autoref{fig:evrard-scaling-gh200} confirms linear scaling of the SPH-part of the application, that is, pair interactions and neighbor list build.
Gravity with Barnes-Hut scales as $\mathcal{O}(n\log n)$ and, with increasing number of subdomains,
exhibits some load imbalance, as the computations for particles closer to the center of the sphere are more expensive. The last category labeled “Tree Build \& Communication” captures
all remaining program parts, i.e., domain decomposition, halo exchanges, and building locally
essential trees, including the computation of multipole moments.

Note that some SPH particle pair interactions are computationally very heavy compared to the previous Lennard-Jones benchmark.
Further, they are not perfectly symmetric, which additionally requires more computations inside each GPU kernel when using the symmetric neighbor list.

In order to obtain a larger statistical sample, in this test, we rebuild the octree and
neighbor list at each time step.
However, in production simulation runs, this would not be not necessary, as the neighbor list update
frequency can be reduced by slightly increasing the search radius, as is typically done in
molecular dynamics simulations.
\begin{figure}[t]
    \centering
    \includegraphics{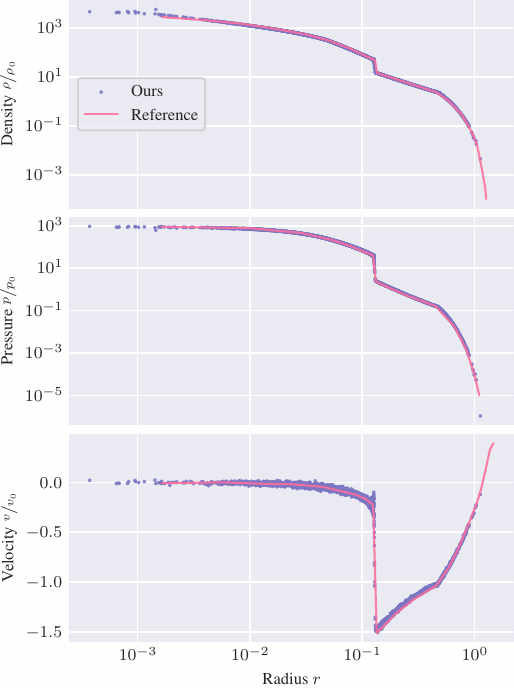}
    \caption{Comparison of our results to the reference data presented in \cite{steinmetz_capabilities_1993}, at $t=0.77$.}
    \label{fig:evrard-validation}
\end{figure}

\begin{figure}[t]
    \centering
    \includegraphics{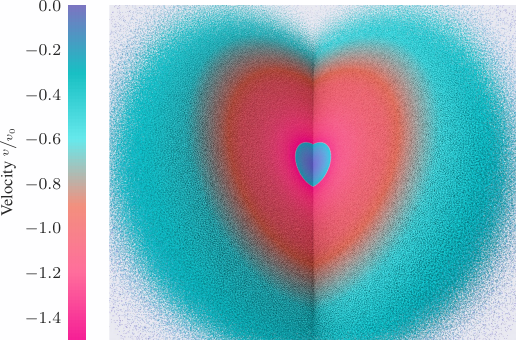}
    \caption{3D visualization of the velocity data at $t=0.77$. One quarter of the sphere is cut out to expose the inside of the sphere.}
    \label{fig:evrard-visulaization}
\end{figure}

To validate our results, we performed an additional simulation with 33.5 million particles and compared the output with the reference values given in \cite{steinmetz_capabilities_1993}
at $t=0.77$.
SPH-EXA implements multiple formulations of SPH; here, we chose the one described in \cite{garcia-senz_conservative_2022}.
The comparison is presented in \autoref{fig:evrard-validation}, and shows excellent agreement with the reference.
Only a small random subset (0.1\,‰) of the 33.5 million particles is plotted in the graph, to maintain PDF responsiveness.
\autoref{fig:evrard-visulaization} shows a 3D visualization of the velocity solution at $t=0.77$.
All particles are rendered here, except for the cut-out quarter of the sphere.

%% file: sec5_limitations.tex
\section{Limitations and Future Work}

The major limitation of our (and other) cluster-based neighbor lists is certainly the overhead introduced by the cluster-cluster interaction computations, as discussed in \autoref{sec:cluster-overhead}.
Especially with smaller neighbor counts, this overhead can be significant.
On modern hardware with large caches and a flexible memory subsystem, such as the NVIDIA GH200, our benchmarks have shown that a full, uncompressed neighbor list may execute faster than clustered approaches even for relatively large neighbor counts, despite several orders of magnitude larger memory footprint and, as a direct consequence, much increased memory transfers.
Only on very high neighbor counts, the clustered neighbor list outperforms the full neighbor list not only in terms of memory footprint, but also run time.
In contrast, on the AMD MI300A GPU, the clustered approach substantially outperforms the full neighbor list in almost all cases, apart from a very small number of neighbors.
Differences in hardware properties and simulation memory requirements thus lead to fundamentally different neighbor list schemes performing optimally.
Tipping points further depend on the specific pair interaction.

It would be interesting to explore the combination of the newly presented compression scheme with a full neighbor list or with clustered neighbor lists with smaller clusters.
This might reduce the cluster overhead with an acceptable memory footprint, due to the SFC compactness guarantees and effective index list compression.

As the compactness guaranty of the Hilbert curve only applies on \emph{average}, it is furthermore possible to create particle distributions with many clusters that do \emph{not} have a compact bounding box.
This mainly introduces overhead when building the neighborhood data structure during octree traversal, but less so when computing the pair interactions, where we will just face more masked-out neighbor clusters.
Besides that, it is also possible to construct many cluster pairs where only very few particle pairs are within the cutoff radius and thus the cluster overhead is much larger than for uniform distributions as presented in \autoref{sec:cluster-overhead}.
However, this limitation also affects grid-based clustering approaches.

%% file: sec6_conclusion.tex
\section{Conclusion}

We have demonstrated that the compactness properties of the Hilbert curve allow us to construct SIMD-friendly clustered neighbor lists as described by Páll and Hess\cite{pall_flexible_2013} without further particle reordering.
In addition, we employed the GPU-native Cornerstone octree library\cite{keller_cornerstone_2023} to build a clustered neighbor list completely on the GPU.
This leads to a highly efficient, GPU-native neighbor list implementation which combines the strengths of
Cornerstone octree, i.e., SFC-based domain decomposition, straightforward integration with multipole methods
for long-range forces, and handling of highly irregular particle distributions and radii, with those of cluster-based pair lists, namely efficient SIMD vectorization of
pair interactions and a small memory footprint.
Finally, we introduce a new neighbor list compression scheme, which further reduces the memory requirements without introducing significant overhead.

Our benchmarks demonstrate good performance and a small memory footprint for a uniform distribution setting as found in molecular dynamics, as well as for an Evrard collapse simulation, a standard test case with
a highly non-uniform particle distribution with variable particle radii that couples hydrodynamical
and gravitational forces.

All code is published under MIT license and available on GitHub\cite{cabezon_sph-exa_2026}.